\documentclass[12pt,final]{article}
\usepackage{amsmath,amssymb,amsfonts,amsthm,slashed,braket}
\usepackage[all]{xy}
\usepackage{mathrsfs}
\usepackage{fancyhdr}
\usepackage{showkeys}

\newcommand\nc{\newcommand}
\nc\linesep{\bigskip}
\nc\newprob[1]{\marginnote{#1}[\parskip]}
\nc\bA{\mathbb A}
\nc\bC{\mathbb C}
\nc\bD{\mathbb D}
\nc\bR{\mathbb R}
\nc\bZ{\mathbb Z}
\nc\bQ{\mathbb Q}
\nc\bP{\mathbb P}
\nc\bV{\mathbb V}
\nc\bW{\mathbb W}
\nc\bG{\mathbb G}
\nc\mf\mathfrak
\nc\mc\mathcal
\nc\mb\mathbb
\nc\brac[1]{\langle#1\rangle}
\nc\abs[1]{\lvert#1\rvert}
\nc\norm[1]{\lVert#1\rVert}
\nc\onto{\twoheadrightarrow}
\nc\into{\hookrightarrow}
\nc\lto{\longrightarrow}
\nc\action{\curvearrowright}
\nc\on\operatorname
\nc\wbar\overline
\nc\what\widehat
\nc\wtilde\widetilde
\nc\nop\DeclareMathOperator
\nc\eps{\varepsilon}
\nc\tsym{\widetilde{\text{Sym}}}
\nc\oarrow[1]{\overset{#1}\to}
\nop\Hom{Hom}
\nop\End{End}
\nop\Aut{Aut}
\nop\im{Im}
\nop\id{id}
\nop\tr{Tr}
\nop\coker{coker}
\nop\Spec{Spec}
\nop\Jac{Jac}
\nop\Ext{Ext}
\nop\Tor{Tor}
\nc\op{\text{op}}
\nop\loc{Loc}
\nop\Frac{Frac}
\nc\ann{\text{ann}}
\nop\QCoh{QCoh}
\nop\Coh{Coh}
\nop\Sym{Sym}
\nop\Hilb{Hilb}
\nop\gr{Gr}
\nop\Tot{Tot}
\nop\Fl{Fl}
\nop\tGamma{\widetilde\Gamma}
\nop\tloc{\widetilde{\text{Loc}}}
\nop\rep{Rep}
\nop\proj{Proj}
\nc\oo[1]{\overset\circ{#1}}
\nop\ospec{\oo{Spec}}
\nop\oTot{\oo{Tot}}
\nop\codim{codim}
\nop\holim{\underset{\lto}{holim}}
\nop\dlim{\underset{\lto}{lim}}

\nop\uHom{\underline{\Hom}}
\nop\dimrel{dim.rel}
\nc\sHom{\mathscr Hom}
\nc\sExt{\mathscr Ext}
\nc\dto{\dashrightarrow}
\nop\rspec{\bf Spec}
\nop\Gal{Gal}
\nop\Ind{Ind}
\nop\Frob{Frob}

\theoremstyle{theorem}

\theoremstyle{remark}

\begin{document} 

\centerline{\large{BPS jumping loci and special cycles}}
\bigskip
\bigskip
\centerline{Shamit Kachru$^1$ and Arnav Tripathy$^2$}
\bigskip
\bigskip
\centerline{$^1$Stanford Institute for Theoretical Physics}
\centerline{Stanford University, Palo Alto, CA 94305, USA}
\medskip
\centerline{$^2$Department of Mathematics, Harvard University}
\centerline{Cambridge, MA 02138, USA}
\bigskip
\bigskip
\begin{abstract}

We study BPS jumping loci, or the subloci in moduli spaces of supersymmetric string vacua where BPS states come into existence discontinuously.  This is a distinct phenomenon from wall-crossing. We argue that these loci should be thought of as special cycles in the sense of Noether-Lefschetz loci or special Shimura subvarieties, which are indeed examples of BPS jumping loci for certain string compactifications. 
We use the Hodge-elliptic genus as an informative tool, suggesting that our work can be extended to understand the jumping behavior of motivic Donaldson-Thomas invariants.

\end{abstract}

\newpage
\tableofcontents

\bigskip
\section{Introduction}

The study of BPS states in supersymmetric string vacua has been intensely pursued for two decades. Motivations have included using BPS spectra as a tool in solving quantum field theories, and finding degeneracies of BPS states at large charge which microscopically match macroscropic predictions for black hole entropy \cite{SW,SV}. Notably, these analyses have usually proceeded by considering various BPS indices in such a way that the resulting count is manifestly moduli-independent, masking the possibility that these indices will detect pair-creation or annihilation of BPS states. Note that such processes are distinct from wall-crossing phenomena, where a wall of marginal stability exists in moduli space with new bound states existing only on one side of the wall.  Here, we are interested in jumping behavior, or higher codimension subloci in the moduli space where extra BPS states appear but such that the spectrum is otherwise unchanged in a neighborhood of the locus.  In other words, we describe the discontinuous variation of the BPS spectrum as one moves in the moduli space.

\medskip
That such behavior should be expected and is even typical is clear from recent work on refined BPS state counts (and is certainly presaged in earlier works such as
\cite{HM,MMS}). Indeed, in \cite{HEG} we study type II string theory compactified on $K3 \times T^2$ and provide a formula for the generating function of $1/4$-BPS states in this $\mc{N} = 4$ vacuum flavored by spin.  The extra flavoring prevents this count from being an index and allows for the observation of pair-creation phenomena. Our generating function is a multiplicative lift of the Hodge-elliptic genus, which we define as 
as a trace in the worldvolume effective string defining the D1-D5 system:
$$Z_{HEG}(q, y, u) = \tr \Big( (-1)^{F_L + F_R} q^{L_0 - c/24} y^{F_L} u^{F_R} \Big)~.$$ 
Here $F_{L,R}$ are the left and right-moving fermion numbers,
and we trace only over states with right-moving part a Ramond ground state. This function notably exhibits jumping behavior at points in moduli space with extra chiral currents, leading to the natural question of sharply characterizing these loci where extra BPS states exist. We study this question in this paper for this and simpler models.

\medskip
In fact, here we distinguish already between two sources of discrepancy between unindexed and indexed counts of BPS states, which we unimaginatively call type 1 and type 2 jumping. By type 1 jumping, we mean the possibility that even the generic value of the unindexed count suffers some discrepancy from the indexed count: in other words, even at a generic point in moduli space, we have boson-fermion cancellations invisible from the point of view of the indexed count. In contrast, by type 2 jumping we mean the possibility of the unindexed count jumping upper semi-continuously as we vary moduli; this upper semi-continuous variation in moduli is exactly parallel to the notion of cohomology jump loci in the mathematics literature, such as in the overview of \cite{BudurWang}. In fact, in many cases the BPS jumping loci exactly reduce to cohomology jump loci in the large-volume limit.\footnote{That this is so follows from the fact that at large volume, BPS states are given by cohomology of D-brane moduli spaces, or cohomology of the moduli spaces of sheaves with fixed Chern classes (in the right duality frame).  The cohomology of these moduli spaces can jump as one varies parameters.}
 In this paper, we mostly focus on type 2 jumping; for the cases we consider, we either have appreciable type 1 jumping, as in 1/4 BPS states in type II toroidal compactifications, or none is expected, as in 1/2 BPS states in heterotic toroidal compactifications.   

\medskip
It is worth emphasizing that we focus on BPS states in theories with highly extended supersymmetry (both on the worldsheet and in space-time).  The story is likely even richer as one moves down in supersymmetry.

\medskip
The next section continues this overview with a general discussion of BPS jumping and a fuller explanation of our notions of type 1 versus type 2 jumping above. We then begin in earnest in section $3$, treating sigma models with toroidal target. Section $4$ increases the mathematical sophistication, treating the case of F-theory on 
elliptic K3 in terms of the classical Noether-Lefshetz theory on the $O(2, 18)$-double coset complex structure moduli space of elliptic K3s.  A very helpful role is played by the dual frame of the heterotic string on $T^2$. We give an analogous story in section $5$ for the $D5$ brane wrapping $K3 \times S^1$ as dual to the heterotic string on $T^4 \times S^1$, focusing on jumping phenomena
in the $O(4, 20)$-double coset Narain moduli space.  In section $6$, we provide a conjectural story for the full D1-D5 system, which extends the moduli space to an $O(4, 21)$-double coset. Finally, we close with a rather speculative discussion in section $7$.  











\section{Generalities on jumping loci}

We first note that the jumping of BPS states is indeed possible, a statement that often sounds odd upon first hearing due to our perhaps over-extensive familiarity with work on protected indices of BPS states. Indeed, if one simply counts the number of BPS states without taking a signed index, one typically expects jumping whenever one has any variation in moduli whatsoever. For example, as we discuss later, even $1/2$-BPS states in $\mc{N} = 4$ supersymmetric models admit plentiful jumping.

\medskip
The above observation also makes it clear that jumping is a very different phenomenon from wall-crossing. In both, the spectrum of BPS states changes discontinuously, but the physical reasoning and types of discontinuity are quite different. In wall-crossing, new bound states appear or disappear due to the energetic favorability of various configurations.  In jumping, on subloci in moduli space (of higher co-dimension), new BPS states simply appear as some particles in the full spectrum of the theory happen to suddenly be annihilated by some supercharge on a closed sublocus. In particular, this last description makes it clear that jumping is an upper semi-continuous phenomenon; in other words, as you tune to a special point in moduli space, more BPS states may suddenly appear and the count may go up, but not down. Of course, the preservation of the topologically-protected index means the new BPS states that appear must appear in boson-fermion pairs.\footnote{This means in boson/fermion pairs relative to the fermion number controlling any fully protected index.  For instance in 4d ${\cal N}=2$ theories, it is vector and hypermultiplets that must pair up as one leaves a jumping locus; the relevant components of those multiplets appear with opposite fermion number parity in relevant worldsheet computations.} 

\medskip
We turn to type 1 versus type 2 jumping as briefly sketched in the introduction. Type 1 jumping denotes the possibility that even at a generic point in moduli, the BPS spectrum exhibits nontrivial bose-fermi cancellations so that the index does not reflect a true count of the BPS states. Detecting this phenomenon should be possible via conformal perturbation theory;
one should be able to see extra BPS states ``pair up'' as one moves away from a symmetric locus in moduli space with a small deformation.
But in fact, on general grounds, we know that type 1 jumping must exist for most classes of models in some sense. Indeed, from black hole entropy considerations, Vafa predicts exponential cancellations of type 1 for all Calabi-Yau threefold compactifications of $M$-theory in \cite{Vjump}. 

\medskip
In the other direction, there are general classes of conjectures predicting the nonexistence of type 1 jumping in various models (for a recent discussion with further references, see \cite{nojump}). For example, in $K3$ compactifications, the indexed count does accurately capture the black hole entropy \cite{SV} and so one may expect no type 1 jumping. Indeed, this expectation is true for the $1/2$-BPS states in this model, but already false for $1/4$-BPS states. As seen in  \cite{Nathan}, there is indeed moderate type 1 jumping (or cancellation) even at generic points in moduli, although not enough to affect asymptotics and thereby (leading behavior in) entropy counts. Another result related to absence of type 1 jumping is Wendland's prediction of the generic decomposition of the Hodge-elliptic genus of the $K3$ $\sigma$-model CFT into $\mc{N} = (4, 4)$ superconformal characters in \cite{Wendland}.  Wendland conjectures the Hodge-elliptic genus of this theory to be the minimal superconformal character compatible with the elliptic genus, with strong associated evidence.

\section{Sigma models with toroidal target}

We begin by explaining the simplest example of the worldsheet theory that governs type II string theory on a single circle $S^1$ of radius $r$. Here, we know the full partition function of the theory, given as\footnote{For a self-contained discussion see \cite{Ginsparg}.} $$Z(S^1_r) = \frac{1}{\eta(\tau) \overline{\eta}(\overline{\tau})} \sum_{m, n} q^{\frac{m}{2r} + nr} \overline{q}^{\frac{m}{2r} - nr}.$$ The analogue of the Hodge-elliptic genus here, without including the extra flavors, is the sum over all pairs $(m, n)$ with vanishing $\overline{q}$-energy, i.e. $$Z_{HEG}(S^1_r) = \frac{1}{\eta(\tau)} \sum_{\frac{m}{2r} = nr} q^{\frac{m}{2r} + nr}.$$ 

\medskip
In this example, the moduli space is determined simply by the compactification radius $r \in \mb{R}^+$ (modulo the T-duality symmetry of $r \mapsto 1/r$), and indeed we see that we have extra contributions beyond the generic answer of $Z_{HEG}(S^1) = \frac{1}{\eta(\tau)}$ precisely when $r^2 \in \mb{Q}^+$, i.e. at some discrete set of subloci inside moduli space, and that at each special sublocus, we have an additional contribution of a rescaled theta function. Indeed, to put this analysis in more fanciful terms, the moduli space of these conformal theories is the double-coset space $$\mc{M} = O(1, 1, \mb{Z}) \backslash O(1, 1, \mb{R}) / (O(1) \times O(1))$$ of lattices of signature $(1, 1)$. This moduli space is parametrizing the location, up to the obvious dualities, of a positive-norm $1$-plane (i.e., a line) in $\mb{R}^{1, 1}$.  The jumping loci occur when new integral lattice vectors are perpendicular to this parametrized $1$-plane. Simple generalizations of this basic pattern will continue to characterize the loci 
exhibiting the emergence of new BPS states in the models we consider here.

\medskip
A very similar story holds for compactification on a torus $T^2$.  The torus is characterized by a complex structure $\tau$ and a 
(complexified) K\"ahler class $\rho$:
$$\tau = \tau_1 + i \tau_2 = {G_{12} \over G_{22}} + i {\sqrt{G} \over G_{22}}
$$
$$\rho = \rho_1 + i \rho_2 = B_{12} + i \sqrt{G}$$
where $G$ is the sigma-model metric, $B$ is the antisymmetric tensor field, and $G = G_{11} G_{22} - G_{12}^2$.

\medskip
One can parametrize the left and right moving momenta in terms of $\tau,\rho$ and momentum and winding numbers $n_1,n_2, m_1, m_2$ as \cite{review}
$$p_L^2 = {1\over 2 \rho_2 \tau_2} \vert (n_1 -\tau n_2) - \rho(m_2 + \tau m_1)\vert^2$$
$$p_R^2 = {1\over 2\rho_2 \tau_2}\vert (n_1 - \tau n_2) -\bar \rho(m_2 + \tau m_1)\vert^2~.$$

\medskip
The duality group consists of $SL(2,{\mathbb Z})$ symmetries acting separately on $\rho$ and $\tau$, together with 
a triplet of ${\mathbb Z}_2$ symmetries:
$$(\tau,\rho) \to (\rho,\tau)$$
$$(\tau,\rho) \to (-\bar\tau,-\bar\rho)$$
$$(\tau,\rho) \to (\tau,-\bar\rho)$$
These can be thought of as mirror symmetry, space-time parity, and world-sheet orientation reversal, respectively.

\medskip
 We can see explicitly from the formulae for $p_{L,R}$ above, that there are loci where there are extra non-trivial
purely left-moving lattice vectors , i.e. vectors with $p_R = 0$.  This happens precisely when
$$\bar\rho = {n_1 - \tau n_2 \over {m_2 + \tau m_1}}$$
for some non-trivial $n_1, n_2, m_1, m_2$.  Given such $n,m$ values, we can scale them by integers keeping $p_R = 0$, 
and therefore we obtain an entire ray of new $p_L$ vectors contributing to the BPS partition function on the jumping locus.

\medskip
A particularly interesting story of deeper mathematical import arises if we restrict to the self-mirror locus, i.e. consider only tori with $\tau=\rho$.
Actually, for our purposes and given the conventions above, it is easier to use the ${\mathbb Z}_2$ symmetries to consider the self-mirror locus as 
$\tau = - \bar\rho$.  The condition $p_R = 0$ then simply becomes
$$m_1 \tau^2 + (m_2 - n_2) \tau + n_1 = 0~.$$
When $\tau$ satisfies a quadratic equation with integer coefficients, the related elliptic curve is said to admit ``complex multiplication."
So we see that on the self-mirror locus, jumping loci line up with curves that admit complex multiplication (and whose mirrors do as well).  These are also loci where the theory
becomes an RCFT, as in \cite{GV}.

\medskip
We conclude this section by noting that the story we saw on the circle and $T^2$ is basically unchanged for general toroidal compactifications.  At least while the moduli space is the double-coset symmetric space for $O(p, p)$, the BPS jumping loci occur first at codimension $p$ and are exactly the loci where new integral lattice vectors emerge orthogonal to the parametrized positive-definite $p$-plane.  The new BPS states which appear are counted by a theta function rescaled by the length of the integral lattice vector. Indeed, we have a fully stratified space, with jumping loci at codimension $p$, further jumping loci at codimension $p$ inside those loci (i.e., at codimension $2p$ inside the total moduli space), and so on, up to a collection of discrete points with maximal BPS spectrum. 

\section{Classical K3 geometry}

We now turn to the study of F-theory on elliptic $K3$ and its dual, the heterotic string on $T^2$.  The moduli space of vacua is $$ \mc{M} = O(2, 18, \mb{Z}) \backslash O(2, 18) / (O(2) \times O(18)).$$ (We ignore here the overall size of the ${\mathbb P}^1$-base in the type IIB description, also known as the heterotic dilaton.) In the heterotic frame, this moduli space is given by the complex and K\"ahler moduli of the compactification $T^2$ together with the 16 Wilson lines for the gauge fields.  The heterotic theory admits an exact computation of the $1/2$ BPS spectrum, by counting Dabholkar-Harvey states \cite{DH}.

\medskip
We can obtain an understanding of the moduli space in terms of $K3$ geometry as follows (here we parallel the discussion in \cite{Greg}).   If $X$ denotes our $K3$ surface, choose a basis of classes $\gamma_I \in H^2($X$,\mathbb{Z})$, $I=1,\dots 22$ in the $K3$ surface such
that
$$\gamma_I \cdot \gamma_J = U^{\oplus 3} \oplus E_8(-1)^{\oplus 2}$$
where
we have the hyperbolic plane $$U = \begin{pmatrix}0&1\\1&0\end{pmatrix}$$
and $E_8(-1)$ denotes the sign-reversed Cartan matrix of $E_8$.

\medskip
The fiber and section (base) of $X$, $F$ and $B$, are algebraic cycles, and hence must be orthogonal to the holomorphic 2-form
$\Omega$ of  $X$.  We can choose 
$$\gamma_1 = F,~~\gamma_2 = B + F~.$$
It is easy to check that their intersection form gives one copy of $U$.

\medskip
Then the complex structure of a general elliptic $K3$ is guaranteed, by the global Torelli theorem, to be expressed as
$$\{ \Omega = \sum_{I=3}^{22} z^J \gamma_J :  \Omega \cdot \bar \Omega > 0,~~\Omega^2 = 0 \}~.$$
Here, we projectively identify
$$z_J \sim \lambda z_J,~~\lambda \in {\mathbb C}^*~.$$

\medskip
To match to the heterotic string, we need to associate to $\Gamma^{2,18} \otimes {\mathbb R}$ a projection onto right and left moving
lattices.  We do this by associating the right-movers to the span of $\Omega, \bar \Omega$, and allowing left-movers to live in the 
orthogonal complement to this span.  In particular, the purely left-moving vectors $\gamma$ are those which have
$$\int_{[\gamma]} \Omega ~=~0.$$
This is precisely the condition that $\gamma$ lives in the Neron-Severi lattice $NS(X)$.  
A generic elliptic $K3$ has ${\rm rank}(NS(X)) = 2$.  However, on special loci, the rank of $NS(X)$ can jump.  On these loci, there
are additional algebraic cycles which can support BPS states in M-theory on $X$ (in the form of wrapped M2-branes).

\medskip
Analyzing the above makes it clear that the BPS jumping loci have a similar description to what we saw previously. Indeed, they occur here first in codimension $2$ and we have a fully stratified structure, with further jumping loci of overall codimension $4$ inside those, and further loci of overall codimension $6$ inside those, and so on up to a discrete countably infinite set of points picked out as the points with locally maximal BPS spectra.  The jumping loci again have purely lattice-theoretic meaning here, tracking exactly where additional integral lattice vectors become purely left-moving. These subloci have an interpretation in terms of the classical moduli space of complex $K3$ surfaces as Noether-Lefschetz loci \cite{GH,rahul}. In classical geometry, the Noether-Lefschetz loci are the subloci in the moduli space of $K3$ tracking jumping of the Picard rank, or the rank of the sublattice of $H_2(K3; \mb{Z})$ that may be realized as linear combinations of curves. In terms of this description, the sublocus of elliptic K3s inside the moduli space of all $K3$s is already a Noether-Lefschetz locus itself, and within it we have further jumping loci labelled by the lattice spanned by additional algebraic curves.

\medskip
For provocation's sake, we also note that in this case, our $O(2, 18)$ double-coset moduli space is a Shimura variety, and the Noether-Lefschetz loci we have identified as BPS jumping loci above are special Shimura subvarieties. We do not explain in detail the mathematical definitions here, instead noting that the Noether-Lefschetz loci are all of the form $$O(2, 20 - \rho, \mb{Z}) \backslash O(2, 20 - \rho) / (O(2) \times O(20 - \rho)),$$ where $\rho$ is the generic Picard rank on the Noether-Lefschetz locus in question. In particular, these loci too have a realization as arithmetic locally symmetric spaces $\Gamma \backslash G / K$, for $\Gamma$ an arithmetic subgroup of some Lie group $G$ with maximal compact subgroup $K$. The notions of Shimura variety and special subvariety both have deep number-theoretic import, and a question we raise now and will return to later is what the analogous distinguishing properties should be for more general supersymmetric vacua (such as 4d $\mc{N} = 2$ theories), when the moduli space is no longer an arithmetic locally symmetric space.

\medskip
We can be more quantitative about the BPS jumping loci discussed above. First, we note that the generic count of BPS states is easily obtained in the heterotic picture as the excited states of $24$ bosonic oscillators, so if $c_n$ is the count of BPS states with energy $n$, we have $$\sum c_n q^{n-1} = \frac{1}{\Delta(\tau)},$$ where $q = e^{2 \pi i \tau}$ and $$\Delta(\tau) = q \prod_{n=1}^{\infty} (1 - q^n)^{24}$$ is the cusp form of weight $12$.  In particular, we have no type 1 jumping of the $1/2$ BPS states in this theory. It remains to understand what the modified count is at the BPS jumping loci, i.e. on the Noether-Lefschetz loci above. As we see from the explicit heterotic computations, if $\Gamma$ is the sublattice of the full $\Gamma^{2, 18}$ that is purely left-moving, the full BPS count is now given 
$$\left(\sum_{p_L \in \Gamma} q^{p_L^2 \over 2}\right) ~{1\over \Delta(\tau)}~.$$
More generally, one could flavor this count by the additional  (left-moving) $U(1)$ quantum numbers that the purely left-moving momentum in $\Gamma$ imparts to the state.\footnote{We are focusing on
states which have $p_R = 0$ in the internal dimensions.  To level match generic such states, one must compactify on an additional circle and choose appropriate winding and momentum.  We focus on these states because we are interested in studying holomorphic BPS
counting functions.}


\medskip

\medskip
To compare to the standard Yau-Zaslow count of curves on $K3$ \cite{YZ}, we should now describe some important differences.  A generic complex K3 has trivial Picard lattice and admits
algebraic curves of no genera.  Yau-Zaslow gets around this by counting curves in all complex structures compatible with a given hyperk\"ahler structure, where the
BPS state count for energy $n-1$ is interpreted as the sum of Gromov-Witten counts for all classes $\beta$ such that $\beta \cdot \beta = 2n-2$.  This correctly
reproduces ${1\over \Delta(\tau)}$.   We provide a more detailed discussion of why the Yau-Zaslow formula admits no jumping in the appendix.

\medskip
In contrast, here we count curves in a fixed complex structure on elliptic $K3$.  The generic elliptic $K3$ has only the algebraic cycles $B$ and $F$.  Following the discussion in \cite{Vjump}, the counting of $M2$-branes on these cycles correctly reproduces ${1\over \Delta}$.  
This is most easily understood from a IIA perspective.  The analogous count considers D2-branes in the class $[B] + n[F]$ for all $n$.  (The single-wrapping of the base corresponds to computing the single heterotic string BPS spectrum).   We can view the $n$ D2 branes wrapping the fibers as having Wilson lines which associate to the points where they live a copy of the Jacobian (the dual torus).  Then the moduli space of the $n$ fiber branes is $n$ copies of the elliptic K3 with dualized fiber, and the BPS state
count will count $H^*(Sym^n(K3))$ at each value of $n$ -- precisely reproducing the desired result.
The enhancements at Noether-Lefschetz loci
in the moduli space of elliptic K3s then contribute additional $\theta$ functions to the BPS count, as described above.\footnote{As noted above, to allow the required $p_L^2$ consistent with level matching in the perturbative heterotic string, one must reduce on an additional circle and add momentum and winding on it.  This is consistent with the dual picture where reduction to M-theory or IIA string theory is needed to do the count.}



\section{Quaternionic Noether-Lefschetz loci}

We are most interested in studying the $D1-D5$ system in the type IIB string on $K3 \times T^2$.  We start slowly in this section, considering the case of one five-brane and no one-branes on $K3 \times S^1$.  Using string dualities, one can map the D5 to a (wrapped) heterotic string in this case.
Then we immediately know the full 1/2 BPS spectrum of interest.  Let us focus on the case where the heterotic compactification is on $T^4 \times S^1$, and
we fix the geometry of the $S^1$.
Then the full moduli space is the double coset $$O(4, 20,\mb{Z}) \backslash O(4, 20) / (O(4) \times O(20)).$$ 

\medskip
Now, from the heterotic string perspective, one has the full even unimodular lattice $\Gamma^{4,20}$, and associated projections onto left and right-moving momenta define the choice of point in the Narain moduli space.  Physical states satisfy
$$N_R - {M^2 \over 4} + {1\over 2}p_R^2 = 0$$
$$N_L - {M^2 \over 4} + {1\over 2} p_L^2   = 0$$
where $N_{R,L}$ are right and left moving oscillator numbers, and $p_R, p_L$ are the projections of the momenta (in the internal space)
into the right and left moving lattices.   For BPS states we require
$$N_R = 0~.$$
Again, we will discuss the counting of BPS states which further have $p_R = 0$, and can satisfy level matching through the compactification on the additional circle
(with appropriate momentum and winding there).

\medskip
We can immediately describe the BPS jumping loci in an analogous way to those of \S4.  The loci are again described by sublattices $\Gamma$ of the $20$-dimensional negative definite sublattice of $\Gamma^{4, 20}$. Explicitly, $$\Gamma^{4, 20} \simeq U^{\oplus 4} \oplus E_8(-1)^{\oplus 2},$$ so its negative-definite sublattice is the sign-reversal of $$I^{\oplus 4} \oplus E_8^{\oplus 2}.$$ The corresponding BPS jumping locus $J_{\Gamma}$ tracks where in the Narain moduli space of lattices $\Gamma$ has become entirely left-moving.  Consequently, if $\Gamma$ has rank $\rho$, the remaining degrees of freedom on this lattice are the remaining $(4, 20 - \rho)$ complement of $\Gamma$ inside the full lattice, and so we have the abstract isomorphism $$J_{\Gamma} \simeq O(4, 20 - \rho, \mb{Z}) \backslash O(4, 20 - \rho) / (O(4) \times O(20 - \rho)).$$ Once again, the jumping loci are arithmetic locally symmetric spaces and arise in all codimensions that are multiples of $4$. More conceptually, these subloci of the hyperk\"ahler moduli space respect its hyperk\"ahler structure, and as such we dub them \textit{quaternionic Noether-Lefschetz loci}.

\subsection{Relation to motivic Donaldson-Thomas}

We pause here to clarify the relation of our work to some of the questions studied in \cite{HEG} -- the characterization of (and jumping of) motivic Donaldson-Thomas invariants of $K3 \times T^2$.  
There, one is interested in BPS states on $K3 \times T^2$ without choosing any additional structure (like a fixed supercharge) -- i.e., one is interested in BPS states annihilated by any supercharge.
But as discussed in \S4, in a theory with extended supersymmetry we are also free to define a count of states that are BPS for some fixed, named supercharge.  We may phrase this more specific count more abstractly by choosing a superconformal subalgebra of the full superconformal algebra of our theory. For example, in the 4d $\mc{N} = 4$ theory given by the $K3 \times T^2$ compactification of the type II string, we could fix an $\mc{N} = 2$ superconformal subalgebra of the full $\mc{N} = 4$ superconformal algebra of the $\sigma$-model with $K3$ target; this determination is the same as choosing a complex structure on our hyperk\"ahler space, or a point in its twistor space. We may then ask about BPS states for this theory considered as an $\mc{N} = 2$ vacuum. We find this question, of counting states annihilated by a fixed supercharge, to often be more natural, but the two questions are naturally related.  This relation is why the jumping loci of this section and the next have some bearing upon the original jumping of motivic Donaldson-Thomas invariants we discussed in \cite{HEG}.

\section{Further generalizations of Noether-Lefschetz loci}

In discussions of AdS/CFT, or of 1/4 BPS black holes in type II compactification on $K3 \times T^2$, it is common to consider the system of $n$ D1-branes and
a single D5-brane.  The low energy theory on the D1-brane is a $\sigma$-model with target $\Hilb^n K3$, and the moduli space (for $n>1$) is the double coset
\cite{Dijkgraaf}
$$O(4,21,{\mathbb Z}) \backslash O(4,21) / (O(4) \times O(21))~.$$

\medskip
The discussion of the previous sections leads us to a natural conjecture for the $1/2$ BPS jumping loci of the $O(4, 21)$ double-coset moduli spaces characterizing the D1-D5 systems; once again, they are quaternionic Noether-Lefschetz loci now given by $O(4, 21 - \rho)$ double-cosets. Indeed, the analysis of the previous section allows us to fully understand the BPS jumping loci intersected with the $O(4, 20)$ double-coset perturbative sublocus.\footnote{Here, we should view the $O(4,20)$ coset as arising from the moduli of the 1 D1-1 D5 system; this system reduces at low energies to a $K3$ $\sigma$-model.  The locus sits in the $O(4,21)$ moduli of the $n$ D1-1 D5 system because on the singular locus
where the target space is literally a symmetric product, the moduli are those of the single $K3$ $\sigma$-model.} This structure certainly lends credence to the hypothesis that the jumping loci in the $O(4, 21)$ double-coset are precisely the analogous Noether-Lefschetz loci.  We now indicate our further expectations for jumping loci in these related systems. 

\medskip
We recall that for one $D1$ and one $D5$ brane, we have the CFT given by the $\sigma$-model to $K3$ with moduli space $$O(4, 20, \mb{Z}) \backslash O(4, 20) / (O(4) \times O(20)),$$ while for any $n > 1$, the system with $n$ $D1$s and one $D5$ is the $\sigma$-model to $\Hilb^n K3$ with moduli space $$O(4, 21, \mb{Z}) \backslash O(4, 21) / (O(4) \times O(21)).$$ 

Focus for a moment on the $O(4,20)$ double coset locus captured by a $K3$ $\sigma$-model.  The $1/2$ BPS states of the $K3$ $\sigma$-model can then be found by considering the dual heterotic picture.
In the heterotic string compacitifed on $T^4 \times S^1$, we can as in \S5\ immediately identify the relevant jumping loci as the quaternionic Noether-Lefschetz loci.
The heterotic string states with distinct momenta on the $S^1$, and arising at the loci $J_{\Gamma}$, translate in a type II picture to $D1-D3-D5$ systems with the number of 
D1 branes related to the $S^1$ momentum, and the D3 branes wrapping curves in the Neron-Severi lattice (which includes $\Gamma$).
Hence, the extent of the jumps at a given jumping locus in a fixed $D1-D3-D5$ system (which sees only a subset of the heterotic string states, as $D1$ number and $D3$ charges are fixed)
will be different than in the heterotic frame.  A fixed D1-D3-D5 system will then see jumping loci which comprise some a priori smaller subset of the quaternionic Noether-Lefschetz loci.  
We defer further discussion of this point.

\medskip
This discussion suggests that the symmetric orbifold locus in moduli space is itself a BPS jumping locus.
We can see this explicitly by checking that the generic answer of \S5 is different from the generic answer in the $O(4, 21)$ double-coset moduli space. One way to see this discrepancy is to do a supergravity computation of the spectrum of BPS states in $AdS_3 \times S^3 \times K3$ supergravity, which should give the answer at a generic point in the $O(4, 21)$ double-coset moduli space, and to note the discrepancy of the answers \cite{Nathan}. This discrepancy certainly shows that the supergravity point is not on the symmetric orbifold locus or any of its U-duality translates (as is well known).  It is possible that it lies on some other jumping locus with a smaller discrepancy from the true generic spectrum, but our physical expectation would be that the field theory dual to large radius $AdS_3$ supergravity has no mysterious extra chiral currents.

\section{Conclusion and speculations}

The main goal of the present paper has been to begin a systematic discussion of ``jumping'' phenomena in capturing the spectrum of BPS states as a function
of moduli in string compactification.  This is a subject of intrinsic interest, it has ties to natural notions in mathematics, and its understanding may help provide characterizations which pin down the exact values (as a function of moduli) of more refined BPS counting functions, like the Hodge-elliptic genus.

\medskip
We pause to reiterate some of the questions we posed in the main body of the text before highlighting some new questions which we believe are of interest. Most broadly, we have an imprecise wish that all moduli spaces of sufficiently supersymmetric string vacua be endowed with some algebraic structure over $\mb{Z}$ or at least some number field.  While this statement as such can not be true on the nose, as for example, $O(4, 20)$ admits no Shimura datum and so its double coset space does not even have natural algebraic structure, we know various work-arounds in many cases. In such terms, we expect the BPS jumping loci to also be arithmetically interesting substacks: when we highlighted this fact before in section $4$, the Noether-Lefschetz loci were special subvarieties in the sense of the theory of Shimura varieties. These special cycles are of deep number-theoretic importance. Indeed, Shimura varieties are already of crucial importance in the Langlands philosophy, giving rise naturally to both automorphic forms and Galois representations. These special cycles play a role in the further investigation of the $L$-functions on both sides: the vanishing of critical values of $L$-functions of automorphic forms is often due to a nonzero period of the form along one of these special cycles, for example. It is worth considering the possibility of such a deep number-theoretic story in more general supersymmetric string vacua.  The beginnings of a parallel automorphic theory over the moduli space of general $\mc{N} = 2$ models has been presaged in a series of interesting works starting with \cite{Yamaguchi}.



\medskip
The study of special points in moduli space picked out in some principled way is of obvious physical interest, and there exists a large literature discussing how to select such special points by means of flux stabilization \cite{flux}, the attractor mechanism \cite{attr}, or rational points in CFT moduli spaces \cite{GV}. We have seen a similar phenomenon taking place in this paper: we have special points characterized by a locally maximal BPS spectrum.  Indeed, it is attractive to conjecture that the points stabilized by the attractor mechanism on a Calabi-Yau threefold are precisely the points with locally maximal BPS spectrum.  To match the moduli dependence (as the attractor mechanism fixes only complex or K\"ahler moduli in IIB or IIA strings, respectively), we mean more precisely that we might expect the attractor points to match the points with locally maximal topological brane spectrum for one of the topological twists, the choice of topological twist corresponding to whether we start with the type IIA or IIB string.  

\medskip
We expect more precise statements to be accessible in models with at least $\mc{N} = 4$ supersymmetry.  For instance, the attractor K3s are those with maximal Neron-Severi lattice \cite{Greg}; these match with the locations in moduli that exhibit maximal jumping of $1/2$ BPS spectra.  We recall Moore's conjectures in \cite{Greg} that the attractor points are defined over number fields and note that, if coupled with our conjecture that attractor points agree with the points of maximal BPS spectrum, suitably interpreted, we find that we expect points of maximal BPS spectrum to be defined over number fields.  We view this statement as the weakest form of our first imprecise conjecture above -- that jumping loci will play a role in the Langlands philosophy in more general, $\mc{N} = 2$
supersymmetric, geometries.

\medskip
One can reasonably ask how our considerations extend to the $1/4$ BPS states in $K3$  $(\times T^2)$ compactification; we make little comment on this extension here other than to note the suggestion in the literature that the $1/4$ BPS state count in type II on $K3\times T^2$ (or its CHL generalizations) should be given by an appropriate Maass-Skoruppa type lift of the $1/2$ BPS state count \cite{JS}.  If one believes that some version of this statement will be true in general, it is reasonable to conjecture that the jumping loci for $1/4$ BPS states are (some subset of) the jumping loci for $1/2$ BPS states. Combining the above several ideas may reasonably lead to the expectation that the jumping loci of the four-variable counting form of  \cite{HEG} are precisely (some subset of) the quaternionic Noether-Lefschetz loci we have defined here.  As this function is (conjecturally) capturing the motivic Donaldson-Thomas invariants of $K3 \times T^2$, this provides an approach to understand jumping phenomena in that context as well.

\medskip
Finally, we note that there are several systems with less supersymmetry where analogues of the phenomena we studied here can be displayed quite concretely.  Two examples should suffice.  In supersymmetric compactification of the heterotic string, the $(0,2)$ worldsheet supersymmetry allows for ``jumping phenomena" where the number of space-time singlet fields enhances at loci in the moduli space.  Examples are visible in Landau-Ginzburg models, for instance in the computations of \cite{KW}.  Similar jumps will occur
in the cohomology problems arising in the computation of supersymmetric brane spectra via matrix factorizations on the Landau-Ginzburg locus \cite{HW}.  It will be interesting to determine the geometry of the jumping loci in such problems (which are surely determinantal varieties, in simple cases) and use them to provide evidence for further, more detailed, conjectures about the distribution of BPS jump loci.

\bigskip
\centerline{\bf{Acknowledgements}}

\medskip
We would like to thank Christoph Keller and Callum Quigley for inspirational talks and Nathan Benjamin, Ying-Hsuan Lin, Natalie Paquette, Roberto Volpato, Edward Witten, Kenny Wong, and Max Zimet for illuminating conversations.  We thank Eric Zaslow for an insightful comment which clarified our first appendix below, and we would especially like to thank Katrin Wendland for an important discussion at the Simons Center in summer 2016. 
The research of S.K. was supported in part by NSF grant
PHY-1316699.

\section*{Appendix: Yau-Zaslow has no jumps}

We explain here why the Yau-Zaslow count \cite{YZ} of $1/2$ BPS states of $D2$ branes wrapping curve classes in $K3$ has no moduli dependence: this count of a subsector of $1/2$ BPS states in an $\mc{N} = 4$ model experiences neither type 1 nor type 2 jumping.  The mathematical description of this problem is that we wish to count curves on a complex $K3$, but as we have fixed no choice of $\mc{N} = 2$ subalgebra of the ${\cal N}=4$ supersymmetry of the $K3$ $\sigma$-model, we allow states which are BPS for any supercharge, not just the fixed supercharges that correspond to the choice of complex structure on the given $K3$.  As such, the count we actually want sums all contributions to holomorphic curve-counts over the full twistor family of the $K3$, varying over all choices of supercharges.  

\medskip
If we instead fixed an $\mc{N} = 2$ subalgebra and merely asked about the curve counts on a fixed complex $K3$, we would indeed have the Noether-Lefschetz jumping loci as more curve classes come into existence.  The claim here is that as we vary over the twistor family, every potential curve class becomes an honest algebraic curve class at precisely one point on the twistor $S^2$.  To see this claim, recall that the twistor family of a hyperk\"ahler space $X$ is described in terms of the K\"ahler form $J \in H^{1, 1}(X)$ and the holomorphic symplectic form $\Omega \in H^{2, 0}(X)$.  Rotating in the twistor family corresponds to rotating $J, \text{Re } \Omega$, and $\text{Im }\Omega$ into one another.  More precisely, we consider an $SU(2)$ action (as the norm $1$ quaternions, say) rotating these three forms into one another, with a $U(1)$ subgroup preserving the complex structure (and merely rotating $\text{Re }\Omega, \text{Im }\Omega$ into one another and so changing the calibration angle of special Lagrangian cycles, this $U(1)$ being the same as the R-symmetry of $\mc{N} = 2$ models).  As such, we identify the twistor space with the homogeneous space $SU(2) / U(1) \simeq S^2$, where the quotient identification is the Hopf fibration.  

\medskip
We seek to understand for which points on the twistor sphere a potential curve class $\gamma \in H_2(X, \mb{Z})$ becomes algebraic.  Recall that a class is algebraic if and only if it is orthogonal to the holomorphic symplectic form $\Omega$, i.e. to its real and imaginary parts.  As these rotate into each other and mix with $J$, we see that the moduli of points on the twistor sphere are always cut down to a constant discrete sublocus unless we are in the degenerate situation where $\gamma$ originally has vanishing pairing with both $\Omega$ and $J$.  However, this eventuality cannot transpire: $\langle \gamma, \Omega \rangle = 0$ implies that $\gamma$ was already an algebraic curve class in the original complex structure, whereupon it must have nonvanishing pairing with $J$ as algebraic classes have positive area (on a smooth surface).  This observation concludes the argument and in fact shows that we have a well-behaved fully-flavored version of the count for the $\mc{N} = 4$ theory as well. 

\medskip
Finally, we expect that similar considerations should explain the moduli independence of the motivic invariants of \cite{KKP}, although further discussion is required to make this expectation precise.

\end{document}